\documentclass[10pt]{article}
\usepackage{amsmath,amssymb,bm,epsf,epsfig,graphicx}
\input epsf.sty
\usepackage[utf8]{inputenc}
\usepackage[english]{babel}
\usepackage{amsmath}
\usepackage{xcolor}
\usepackage{latexsym} 
\usepackage{amsfonts}
\usepackage{mathtools}
\usepackage{amssymb}
\usepackage[left=3cm,right=3cm,top=3.1cm,bottom=3.1cm]{geometry}
\usepackage{cite}

\usepackage{amsthm}
\newtheorem{claim}{Claim}

\usepackage{hyperref}
\numberwithin{equation}{section}

\def\nubos{\nu_\mathrm{boson}}
\def\nusuper{\nu_\mathrm{super}}

\def \be {\begin{eqnarray}}
\def \ee {\end{eqnarray}}
\def \bdm {\begin{displaymath}}
\def \edm {\end{displaymath}}

\def\0{\nonumber}

\def\bS{\overline\Sigma}
\def\bs{\overline\sigma}
\def\S{\Sigma}
\def\s{\sigma}

\begin{document}

\vspace*{3.1cm}

\centerline{\Large \bf Taming boundary condition changing operator anomalies }\vspace{.4cm}
\centerline{\Large \bf with the tachyon vacuum} \vspace{.2cm}
\vspace*{.1cm}

\begin{center}

{\large Theodore Erler${^{(a)}}$\footnote{Email: tchovi at gmail.com}, Carlo Maccaferri$^{(b),(d)}$\footnote{Email: maccafer at gmail.com},  Ruggero Noris$^{(c),(d)}$\footnote{Email: ruggero.noris at polito.it} } 
\vskip .7 cm
$^{(a)}${\it Institute of Physics of the ASCR, v.v.i.}\\
{\it Na Slovance 2, 182 21 Prague 8, Czech Republic}\\
\vskip .5 cm
$^{(b)}${\it Dipartimento di Fisica, Universit\'a di Torino}, \\
{\it Via Pietro Giuria 1, I-10125 Torino, Italy}\\
\vskip .5 cm
$^{(c)}${\it DISAT, Politecnico di Torino,},\\
{\it C.so Duca degli Abruzzi 24, I-10129 Torino, Italy}\\
\vskip .5 cm
$^{(d)}${\it INFN  Sezione di Torino and Arnold-Regge Center}\\
{\it Via Pietro Giuria 1, I-10125 Torino, Italy}\\

\end{center}

\vspace*{6.0ex}

\centerline{\bf Abstract}
\bigskip
Motivated by the appearance of associativity anomalies in the context of superstring field theory, we give a generalized solution built from boundary condition changing operators which can be associated to a generic tachyon vacuum in the $KBc$ subalgebra of the Okawa form. We articulate sufficient conditions on the choice of tachyon vacuum to ensure that ambiguous products do not appear in the equations of motion. 

 \vfill \eject

\baselineskip=16pt

\tableofcontents
\section{Introduction}\label{intro}
In recent years a remarkably simple solution to the equation of motion of open bosonic string field theory has been found \cite{EM}, which can relate any pair of time independent open string backgrounds sharing the same closed string bulk\footnote{The solution of \cite{EM} is a subtle but important refinement of the solution found some time ago by Kiermaier, Okawa, and Soler \cite{KOS}. Various applications have been studied in \cite{Ishibashi:2018ynb, Kishimoto:2014yea, Maccaferri:2015cha, Ishibashi:2016xak}}. The solution is defined by a choice of tachyon vacuum $\Psi_{\rm tv}$ and a pair of string fields $(\S, \bS)$,  which change the worldsheet boundary conditions from the starting background BCFT$_0$ defining the reference open string field theory, to the target background BCFT$_*$ we wish to describe. The solution takes the form
\be
\Psi=\Psi_{\rm tv}- \S\Psi_{\rm tv}\bS,\label{EM}
\ee
and the equations of motion are satisfied provided 
\be
&\ &\!\!\!\!\!\!\!\!\!\!\!\!\!\!\!\!\!\!\!\!\! Q_{\Psi_{\rm tv}}\S=Q_{\Psi_{\rm tv}}\bS=0,\\
&\ & \!\!\!\! \bS\,\S=1.\label{bSS}
\ee
The objects $(\S, \bS)$ can be expressed in terms of string fields $(\s,\bs)$ representing insertions of weight zero primary boundary condition changing operators in correlation functions on the cylinder \cite{EM}. They multiply as
\be
\bs\s&=&1,\label{gg1}\\
\s\bs&=&\frac{g_*}{g_0},\label{gg2}
\ee
where $g$ is the disk partition function of the corresponding BCFT
\be
g_x\equiv\langle 1\rangle^{{\rm BCFT}_x}_{disk}.
\ee
The first relation $\bs\s=1$ is the one which is needed to realize \eqref{bSS}, but the second creates potential problems with associativity  and renders the triple products $\s\bs\s$  undefined if, as is typically the case, $g_0\neq g_*$, $i.e.$ if the initial and final D-branes systems have a different mass. However, this ambiguity does not appear in essential computations with the solution.

After \cite{EM} it was immediately clear that the solution could be generalized to superstring field theory, at least in its formal structure.  However, there is an unexpected setback: this time, ambiguous products of boundary condition changing operators appear explicitly in the equations of motion. For simplicity we discuss the solution to the Chern-Simons-like equations of motion of cubic superstring field theory at picture zero \cite{PTY,Russians}, but analogous considerations apply in the Wess-Zumino-Witten-like formulation \cite{Berkovits}. The solution of the superstring contains a term  
\begin{equation}\frac1{\sqrt{1+K}}\,\s\bs \,B\gamma^2\frac1{\sqrt{1+K}},\end{equation}
where the unwanted product $\s\bs=\frac{g_*}{g_0}$ explicitly appears. We propose the following resolution to this problem. In \cite{EM} the state $\Psi_{\rm tv}$ was assumed to be the  ``simple" tachyon vacuum of \cite{simple-tv}:
\begin{equation}
\Psi_{\rm tv}=\frac{1}{\sqrt{1+K}}\Big(c+Q(Bc)\Big)\frac{1}{\sqrt{1+K}}.
\end{equation} 
The factors $\frac{1}{\sqrt{1+K}}$ are too similar to the identity string field, and for the superstring do not provide sufficient separation between the boundary condition changing operators to avoid ambiguous products. On the other hand, if the state $\Psi_{\rm tv}$ had been Schnabl's solution \cite{Schnabl},
\begin{equation}
\Psi_{\rm tv}=\sqrt{\Omega}cB\frac{K\Omega}{1-\Omega}c\sqrt{\Omega}+\sqrt{\Omega}Q(Bc)\sqrt{\Omega},
\end{equation}
the factors $\sqrt{\Omega}$ would ensure that $\s$ and $\bs$ are always separated by a surface of nonzero width, and ambiguous products cannot appear. 

The above remedy is quite simple. But the technically interesting and nontrivial question is ``how much" $\s$ and $\bs$ need to be separated to avoid ambiguities. Apparently, the simple tachyon vacuum does not provide enough separation, while at the other extreme Schnabl's tachyon vacuum probably gives more than necessary. Addressing this question requires formulating a sufficient criterion for the absence of short-distance ambiguities in expressions involving products of boundary condition changing operators and elements of the wedge algebra. This understanding is likely to be useful not only in the present work but for evaluating other computations related to the solution of \cite{EM} and possible generalizations.

This paper is organized as follows. In section \ref{sec2} we describe the superstring generalization of the solution of \cite{EM}, this time allowing for the possibility that $\Psi_{\rm tv}$ could be a generic tachyon vacuum of the Okawa form \cite{Okawa}. We give two formally equivalent expressions for the solution, differing by whether or not BRST variations of boundary condition changing operators are explicitly evaluated. In section \ref{sec3} we consider short-distance singularities in the solution, in particular those which concern collisions of two or three boundary condition changing operators, leading respectively to OPE divergences or associativity anomalies. In this analysis the dual $\mathcal{L}^-$ level expansion \cite{dual-L} is helpful for making precise statements. The headline conclusions are as follows: First, if the solution is expressed in the form where BRST variations of the boundary condition changing operators are not explicitly evaluated, OPE divergences can appear in individual terms which, however, cancel. Second, if $\Psi_{\rm tv}$ provides even a little more separation between boundary condition changing operators than the simple tachyon vacuum, there are no ambiguous products in the superstring equations of motion. In section \ref{sec4} we observe that solutions for different choices of $\Psi_{\rm tv}$ can be related by automorphisms reflecting reparameterization symmetries of the spectrum of $K$. We end with concluding remarks.

\section{Solution}\label{sec2}

We consider the solution to the Chern-Simons-like equations of motion for the superstring. The bosonic string solution can be obtained by setting $\gamma$ ghosts to zero. The solution lives in a subalgebra of states given by multiplying the string fields
\begin{equation}K,\ \ B,\ \ c,\ \ \gamma^2,\ \ \s,\ \ \bs,\ \ Q\s,\ \ Q\bs,\end{equation}
which represent operator insertions on the identity string field. Our conventions for  $K,B,c,\gamma^2$ follow~\cite{exotic}, to which we refer the reader for definitions and algebraic relations (see also section \ref{sec4}). The fields $\s$ and $\bs$ are defined as in \cite{EM}, but with the additional specification (for the superstring) that they represent insertions of matter superconformal primaries of dimension $0$. This implies that their BRST variations are given by 
\begin{eqnarray}
Q\s & = & c\partial \s+\gamma \delta \s,\label{Qs}\\
Q\bs &=& c\partial\bs+\gamma\delta\bs,\label{Qbs}
\end{eqnarray}
where
\begin{equation}\delta\s = G_{-1/2}\cdot \s\end{equation}
represents the worldsheet supersymmetry variation. For the superstring, $Q\s$ and $Q\bs$ cannot be expressed using $K,B,c,\gamma^2,\s,\bs$ and are independent generators of the subalgebra.

We consider a class of tachyon vacuum solutions of the Okawa form \cite{Okawa,Erler-tv-cubic}\footnote{The most general tachyon vacuum solution in the $KBc$ subalgebra is discussed in \cite{Jokel}.}
\begin{equation}\Psi_{\rm tv} = \sqrt{F}\left(c\frac{B}{H}c + B\gamma^2\right)\sqrt{F},\label{tvs}\end{equation}
where $F=F(K)$ is a suitably well-behaved element of the wedge algebra (a real function of $K$) satisfying the conditions \cite{exotic,SSFII,MurataSchnabl}
\begin{equation}F(0)=1,\ \ F'(0)<0,\ \ F(\infty)= 0,\ \ F(K)<1.\label{F_cond}\end{equation}
We introduce $H=H(K)$ which is related to $F$ through
\begin{equation}H = \frac{1-F}{K}.\end{equation}
The simple tachyon vacuum of \cite{simple-tv} is defined by equating
\begin{equation}
H \equiv F\ \ \to\ \ F = \frac{1}{1+K}\ \ \ \ ({\rm simple\ tachyon\ vacuum}).
\end{equation}
This choice of tachyon vacuum was assumed in \cite{EM}. Presently we are concerned with more general choices of $F$.

In the following we will not need to be specific about the choice of $F(K)$, but nevertheless it may be helpful to describe a representative class of tachyon vacuum solutions which are sufficient for our analysis. Consider a one-parameter family of $F(K)$s of the form
\begin{equation}F(K) = \left(1-\frac{1}{\nu}K\right)^\nu = \frac{(-\nu)^{-\nu}}{\Gamma(-\nu)}\int_0^\infty dt\, t^{-\nu-1} e^{\nu t}\Omega^{t},\label{F_ex}\end{equation}
The parameter $\nu<0$ represents the leading level of $F(K)$ and its tachyon vacuum solution in the dual $\mathcal{L}^-$ level expansion \cite{dual-L}. Note that as $\nu$ becomes increasingly negative, the contribution from states close to the identity string field in the integrand becomes further suppressed. The case $\nu=-1$ corresponds to the simple tachyon vacuum, and $\nu=-\infty$ corresponds to Schnabl's solution. One can show that
\begin{eqnarray}
H(K) &=& \int_0^\infty dt\, \frac{\Gamma(-\nu,-\nu t)}{\Gamma(-\nu)}\Omega^t, \\
\frac{F(K)}{H(K)} &=& \int_0^\infty dt\left[e^{\nu t}\left(\nu+\frac{d}{dt}\right)\frac{d}{dt}E_{-\nu}\Big((-\nu t)^{-\nu}\Big)\right]\Omega^t,\ \ \ \ (\nu< -1),\end{eqnarray}
where $\Gamma(s,x)$ is the (upper) incomplete gamma function and $E_\alpha(x)$ is the Mittag-Leffler function. Therefore the solution can be  written explicitly as a multi-dimensional integral over wedge states with insertions. The singularities of $H$ and $F/H$ in the negative half of the complex plane imply that the inverse Laplace transforms fall off exponentially for large $t$:
\begin{equation}\frac{\Gamma(-\nu,-\nu t)}{\Gamma(-\nu)}\sim e^{\nu t},\ \ \ \ \ \ \ \ e^{\nu t}\left(\nu+\frac{d}{dt}\right)\frac{d}{dt}E_{-\nu}\Big((-\nu t)^{-\nu}\Big)\sim e^{\nu(1-\cos\frac{2\pi}{\nu})t}.\end{equation}
Therefore, unlike for Schnabl's solution, for finite $\nu<0$ we do not need to place a cutoff on the upper limit of integration over wedge states and subtract a phantom term. 

Once we have $\Psi_{\rm tv}$ we can build the solution 
\begin{equation}\Psi = \Psi_{\rm tv} - \S\Psi_{\rm tv}\bS,\label{bcc}\end{equation}
where $\S$ and $\bS$ are given by
\begin{eqnarray}
\S & =& Q_{\Psi_{\rm tv}}(\sqrt{H}\s B\sqrt{H}),\\
\bS & =& Q_{\Psi_{\rm tv}}(\sqrt{H}B\bs\sqrt{H}).
\end{eqnarray}
One may confirm that 
\begin{equation}
Q_{\Psi_{\rm tv}}\S = Q_{\Psi_{\rm tv}}\bS = 0,\ \ \ \bS\S =1,
\end{equation}
so that \eqref{bcc} formally satisfies the equations of motion. For later analysis it will be useful to substitute the definitions and expand the solution explicitly. We give the solution in two forms. In the first form, we leave $Q\s$ and $Q\bs$ as they are, instead of substituting \eqref{Qs} and \eqref{Qbs}. We then have
\begin{eqnarray}
\S &=&\sqrt{H}\s\frac{1}{\sqrt{H}} + \sqrt{H}Q\s B\sqrt{H}+\sqrt{H}\left[\sqrt{\frac{F}{H}}c\sqrt{\frac{F}{H}},\s\right]B\sqrt{H},\\
\bS &=& \frac{1}{\sqrt{H}}\bs\sqrt{H} - \sqrt{H}B Q\bs\sqrt{H}+\sqrt{H}B\left[\bs,\sqrt{\frac{F}{H}}c\sqrt{\frac{F}{H}}\right]\sqrt{H},\label{SbSQ}
\end{eqnarray}
and
\begin{eqnarray}
\Psi &=& \sqrt{F}\left(c\frac{B}{H}c+B\gamma^2\right)\sqrt{F}-\sqrt{H}\s \sqrt{\frac{F}{H}}\left(c\frac{B}{H}c+B\gamma^2\right)\sqrt{\frac{F}{H}}\bs \sqrt{H}+\sqrt{H}Q\s B FQ\bs\sqrt{H}\nonumber\\
&\ & -\left(\sqrt{H} Q\s B\sqrt{\frac{F}{H}}c\sqrt{\frac{F}{H}}\bs \sqrt{H} +{\rm conj.}\right)-\left(\sqrt{H}\left[\sqrt{\frac{F}{H}} c\sqrt{\frac{F}{H}},\s\right]B\sqrt{\frac{F}{H}}c\sqrt{\frac{F}{H}}\bs\sqrt{H}+{\rm conj.}\right)\nonumber\\
&\ &-\left(\sqrt{H}Q\s BF\left[\bs,\sqrt{\frac{F}{H}}c\sqrt{\frac{F}{H}}\right]\sqrt{H}+{\rm conj.}\right)-\sqrt{H}\left[\sqrt{\frac{F}{H}} c\sqrt{\frac{F}{H}},\s\right]BF\left[\bs,\sqrt{\frac{F}{H}}c\sqrt{\frac{F}{H}}\right]\sqrt{H},\nonumber\\\label{solutionQ}
\end{eqnarray}
where ``conj" denotes the reality conjugate of the previous term (see \eqref{real_rel}). In the second form of the solution, we expand the BRST variations using \eqref{Qs} and \eqref{Qbs}:
\begin{eqnarray}
\S &=& \sqrt{H}Bc\s\frac{1}{\sqrt{H}} +\sqrt{H}cB\frac{1}{H}\s\sqrt{H}-\sqrt{H}\left[\left[c,\sqrt{\frac{F}{H}}\right] \sqrt{\frac{F}{H}},\s\right]B\sqrt{H} + \sqrt{H}\gamma\delta \s B\sqrt{H},\ \ \ \ \ \\
\bS &=& \frac{1}{\sqrt{H}} \bs cB\sqrt{H} +\sqrt{H}\bs\frac{1}{H}Bc\sqrt{H}-\sqrt{H}B\left[\bs,\sqrt{\frac{F}{H}}\left[\sqrt{\frac{F}{H}},c\right]\right]\sqrt{H} - \sqrt{H}B \gamma\delta \bs \sqrt{H}.\ \ \ \ 
\end{eqnarray}
The solution becomes:
\begin{eqnarray}
\Psi \!\!\!\!&=\!\!\!\!& \sqrt{F}\left(c\frac{B}{H}c+B\gamma^2\right)\sqrt{F}-\sqrt{H}c\frac{1}{H}\s B F \bs\frac{1}{H}c\sqrt{H} -\sqrt{H}\s\sqrt{\frac{F}{H}}B\gamma^2\sqrt{\frac{F}{H}}\bs\sqrt{H}\nonumber\\
&\ & +\sqrt{H}\gamma\delta\s BF\gamma\delta\bs\sqrt{H}-\left(\sqrt{H}\gamma\delta\s BF\bs \frac{1}{H}c\sqrt{H}+{\rm conj.}\right) -\sqrt{H}\s\left[\sqrt{\frac{F}{H}},c\right]\frac{B}{H}\left[c,\sqrt{\frac{F}{H}}\right]\bs\sqrt{H}\nonumber\\
&\ & +\left(\sqrt{H}c\frac{B}{H}\s\sqrt{\frac{F}{H}}\left[\sqrt{\frac{F}{H}},c\right]\bs\sqrt{H}+{\rm conj.}\right)-\left(\sqrt{H}\gamma\delta\s B\sqrt{\frac{F}{H}}\left[c,\sqrt{\frac{F}{H}}\right]\bs\sqrt{H}+{\rm conj.}\right)\nonumber\\
&\ & +\left(\!\sqrt{H}c\frac{B}{H}\s F \left[\bs,\sqrt{\frac{F}{H}}\left[\sqrt{\frac{F}{H}},c\right]\right]\sqrt{H}\!+\!{\rm conj.}\!\right)\!+\!\left(\!\sqrt{H}\gamma\delta\s BF \left[\bs,\sqrt{\frac{F}{H}}\left[\sqrt{\frac{F}{H}},c\right]\right]\sqrt{H}\!+\!{\rm conj.}\!\right)\nonumber\\
&\ &+\left(\sqrt{H}\s \left[\sqrt{\frac{F}{H}},c\right]B\sqrt{\frac{F}{H}}\left[\bs,\sqrt{\frac{F}{H}}\left[\sqrt{\frac{F}{H}},c\right]\right]\sqrt{H}+{\rm conj.}\right)\nonumber\\
&\ &-\sqrt{H}\left[\left[c,\sqrt{\frac{F}{H}}\right]\sqrt{\frac{F}{H}},\s\right]B F\left[\bs,\sqrt{\frac{F}{H}}\left[\sqrt{\frac{F}{H}},c\right]\right]\sqrt{H}.
\label{solutionnoQ}\end{eqnarray}
This form of the solution generalizes the expression given in \cite{EM}. If we set the $\gamma$ ghosts to zero and set $H=F$, only the first two terms survive, giving
\begin{equation}\Psi = \frac{1}{\sqrt{1+K}}cB(1+K)c\frac{1}{\sqrt{1+K}}-\frac{1}{\sqrt{1+K}}c(1+K)\s\frac{B}{1+K}\bs(1+K)c\frac{1}{\sqrt{1+K}},\end{equation}
in agreement with \cite{EM}. On the other hand, we can do the same thing for the solution as expressed in \eqref{solutionQ}. This gives
\begin{eqnarray}\Psi &=&  \frac{1}{\sqrt{1+K}}c(1+K)Bc\frac{1}{\sqrt{1+K}}-\frac{1}{\sqrt{1+K}}\s c(1+K)B c\bs \frac{1}{\sqrt{1+K}}\nonumber\\
&\ &+\frac{1}{\sqrt{1+K}}Q\s  \frac{B}{1+K}Q\bs\frac{1}{\sqrt{1+K}} -\frac{1}{\sqrt{1+K}} (Q\s B c\bs-\s c BQ\bs)\frac{1}{ \sqrt{1+K}}.\end{eqnarray}
This expression for the solution, which does not appear in \cite{EM}, has a potentially problematic collision between $\s$ and $\bs$ in every term besides the first. We discuss this more in the next section. 

\section{Taming Anomalies}\label{sec3}

We wish to determine sufficient conditions on the choice of tachyon vacuum, or equivalently $F(K)$, such that the solution suffers no difficulties from collisions of boundary condition changing operators. There can be problems if $F(K)$ is too similar to the identity string field, so that there is not ``enough surface" to prevent $\s$ from colliding with $\bs$. The degree of similarity to the identity string field can be quantified by the rate of decay of $F(K)$ as $K\to\infty$ \cite{dual-L}. For definiteness we assume that it decays as a power:
\begin{equation}F(K)\sim K^\nu,\ \ \ K\to\infty,\end{equation}
where $\nu$ is a real number less than zero. The class of $F(K)$s described in \eqref{F_ex} show precisely this asymptotic behavior for large $K$. We have
 \begin{equation}H\sim\frac{1}{K},\ \ \ \sqrt{\frac{F}{H}}\sim K^{\frac{\nu+1}{2}},\ \ \ K\to\infty.\end{equation}
The more quickly $F(K)$ vanishes as $K\to\infty$, the less ``identity-like" the tachyon vacuum becomes, and the more regular the solution should appear from the point of view of collisions of boundary condition changing operators. Since the bounds we derive in the bosonic and supersymmetric cases are different, we use $\nubos$ to indicate the rate of decay of $F(K)$ for the bosonic string solution and $\nusuper$ for the superstring solution. The simple tachyon vacuum corresponds to $\nu=-1$, which for the superstring already poses difficulties. Therefore $\nusuper$ should be bounded from above  by $-1$:
\begin{equation}\nusuper<-1.\end{equation}
The question is whether this bound is sufficient, or should be further strengthened.

Associativity anomalies are related to collisions of three boundary condition changing operators, but there can already be problems from the collision of two. For $\s$ and $\bs$ themselves the OPE is regular,
\begin{equation}\s(s)\bs(0)={\rm regular},\end{equation}
but this still allows singularities in OPEs with $\partial \s$ and $\partial \bs$:
\begin{eqnarray}
\s(s)\partial\bs(0)&\sim & {\rm\ less\ singular\ than\ simple\ pole},\nonumber\\
\partial\s(s)\partial\bs(0)&\sim & {\rm\ less\ singular\ than\ double\ pole}.\label{sdsOPE}
\end{eqnarray}
Such singularities can appear when a relevant field is found in the OPE between $\s$ and $\bs$ \cite{EM}. For the superstring we additionally assume
\begin{equation}\s(s)\delta\bs(0)={\rm regular}.\end{equation}
Together with \eqref{sdsOPE} this implies
\begin{eqnarray}
\delta\s(s)\delta\bs(0)&\sim & {\rm\ less\ singular\ than\ simple\ pole},\nonumber\\
\partial\s(s)\delta\bs(0)&\sim & {\rm\ less\ singular\ than\ simple\ pole}.
\end{eqnarray}
We make the following claim:
\begin{claim}Let  $\mathcal{O}_{1}$ represent $\s,\partial\s$ or $\delta\s$ and $\mathcal{O}_2$ represent $\bs,\partial\bs$ or $\delta\bs$. Then the state 
\begin{equation}\mathcal{O}_1 G(K)\mathcal{O}_2\end{equation}
suffers from no OPE divergence provided that its leading level in the dual $\mathcal{L}^-$ level expansion \cite{dual-L} is less than or equal to $0$ if the state is GSO even, and less than or equal to $1/2$ if the state is GSO odd. \label{claim1}
\end{claim}
\noindent This is a technical way of saying that if we expand $G(K)$ as an integral over wedge states
\begin{equation}\mathcal{O}_1 G(K)\mathcal{O}_2 =\int_0^\infty dt\, g(t) \mathcal{O}_1\Omega^t\mathcal{O}_2,\end{equation}
any singularity which appears in the integrand towards $t=0$ must be integrable. Consider the solution expressed in terms of $Q\s$ and $Q\bs$ as written in \eqref{solutionQ}. The solution contains the terms
\begin{eqnarray}
& &\sqrt{H}Q\s B FQ\bs\sqrt{H}, \nonumber\\
& &\!\!\!\!\!\!\!\! \sqrt{H} Q\s B\sqrt{\frac{F}{H}}c\sqrt{\frac{F}{H}}\bs \sqrt{H}. 
\end{eqnarray}
Computing the BRST variations and ignoring ghosts, which in this case are unimportant, the matter sector component of these states contains the respective factors:
\begin{eqnarray}
\partial\s F\partial \bs &\sim & \partial\s K^\nu\partial\bs, \\
\partial\s \frac{F}{H}\bs &\sim & \partial\s K^{\nu+1}\bs, \ \ \ K\to\infty.
\end{eqnarray} 
By claim \ref{claim1}, these states do not suffer OPE divergence if 
\begin{equation}\nu\leq -2\ \ \ {\rm (no\ OPE\ divergences\ in}\ \eqref{solutionQ}).\end{equation}
The remaining terms do not alter this bound. Therefore, $F(K)$ must fall off as $K^{-2}$ or faster to be certain that OPE divergences are absent from  \eqref{solutionQ}, in either bosonic or supersymmetric cases. However, the original solution of \cite{EM} is finite, even though this bound is violated. In this case \eqref{solutionQ} should be seen as a singular representation of an otherwise regular solution; its OPE divergences formally cancel. This is made manifest in the second form of the solution \eqref{solutionnoQ}, which is much safer from OPE divergence:
\begin{equation}\nubos\leq 0\ \ \rm{or}\ \ \nusuper\leq -1\ \ \ {\rm (no\ OPE\ divergences\ in}\ \eqref{solutionnoQ}).\end{equation}
Therefore when $-2<\nubos\leq0$ or $-2<\nusuper\leq-1$ the solution as written in \eqref{solutionnoQ} will be free of OPE divergence, but divergences may still be present in \eqref{solutionQ}. 

Now let's turn to issues which concern three boundary condition changing operators. These do not affect the solution $\Psi$ by itself as a state, since it only contains two boundary condition changing operators. However, they concern the validity of the equations of motion
\begin{equation}Q\Psi + \Psi^2 =0,\end{equation}
since $\Psi^2$ contains four boundary condition changing operators. We make the following claim:
\begin{claim} Let $\mathcal{O}_1,\mathcal{O}_2$ and $\mathcal{O}_3$ represent three primary operators, and consider the state
\begin{equation}\mathcal{O}_1G_1(K)\mathcal{O}_2G_2(K)\mathcal{O}_3.\end{equation}
Simultanous collision of all three operators do not render this state undefined provided that its leading level in the dual $\mathcal{L}^-$ level expansion is less than $h$, where $h$ is the lowest dimension of a primary operator which has nonvanishing contraction with the state.\footnote{There is a straightforward generalization concerning products of $n$ primary operators with wedge states. The case $n=2$ almost implies claim \ref{claim1}. However, claim \ref{claim1} is slightly stronger, since the OPEs of the boundary condition changing operators are more regular that would be implied by conformal invariance alone. }\label{claim2}
\end{claim}
\noindent To understand this claim, we contract with a test state
\begin{equation}\Omega\mathcal{O}\Omega^\infty,\label{test}\end{equation}
where $\mathcal{O}$ is a primary operator. Since the singularity which interests us concerns short distance behavior when $\mathcal{O}_1,\mathcal{O}_2,\mathcal{O}_3$ collide, the precise form of the test state is not crucial. We choose \eqref{test} since the sliver state allows us to bypass a conformal transformation from the cylinder to the upper half plane which complicates the computation without changing the result. Therefore we consider the overlap
\begin{equation}
{\rm Tr}\Big[\Omega^\infty\mathcal{O}_1 G_1(K)\mathcal{O}_2G_2(K)\mathcal{O}_3\Omega\mathcal{O}\Big]=\int_0^\infty dt_1dt_2\,g_1(t_1)g_2(t_2)\Big\langle\mathcal{O}_1(t_1+t_2)\mathcal{O}_2(t_2)\mathcal{O}_3(0)\mathcal{O}(-1)\Big\rangle_{\rm UHP},
\end{equation}
where $g_1$ and $g_2$ are the inverse Laplace transforms of $G_1$ and $G_2$. We change the integration variables,
\begin{equation}
L=t_1+t_2,\ \ \ \ \theta = \frac{t_2}{t_1+t_2},
\end{equation}
and apply a conformal transformation to the 4-point function so that $\mathcal{O}_1$ is inserted at $1$, $\mathcal{O}_3$ is inserted at $0$, and $\mathcal{O}$ is inserted at infinity. This gives
\begin{eqnarray}
&\ &\!\!\!\!\!\!\!\!\!\!\!\!\!\!\!\!\!\!\!\!
{\rm Tr}\Big[\Omega^\infty\mathcal{O}_1 G_1(K)\mathcal{O}_2G_2(K)\mathcal{O}_3\Omega\mathcal{O}\Big]\nonumber\\
&=&\int_0^\infty dL\int_0^1 d\theta\, L g_1(L(1-\theta))g_2(L\theta)\left(\frac{1}{L}\frac{1}{L+1}\right)^{h_1}\left(\frac{1}{L}\frac{L+1}{(L\theta+1)^2}\right)^{h_2}\left(\frac{L+1}{L}\right)^{h_3}\left(\frac{L}{L+1}\right)^{h}\nonumber\\
&\ & \times\left\langle\mathcal{O}_1(1)\mathcal{O}_2\left(\frac{L+1}{L\theta+1}\theta\right)\mathcal{O}_3(0)I\circ\mathcal{O}(0)\right\rangle_{\rm UHP},\label{argclaim2}
\end{eqnarray}
where $I(z)=-1/z$ is the BPZ conformal map. We are interested in the behavior of the integrand towards $L=0$, which is when $\mathcal{O}_1,\mathcal{O}_2$ and $\mathcal{O}_3$ collide. For small $L$ we have
\begin{equation}g_1(L(1-\theta))\sim L^{-\nu_1-1}(1-\theta)^{-\nu_1-1},\ \ \ g_2(L\theta)\sim L^{-\nu_2-1}\theta^{-\nu_2-1}\ \ \ \ \ {\rm (small\ } L),\label{g1g2}\end{equation}
where $\nu_1,\nu_2$ are the leading levels of the dual $\mathcal{L}^-$ expansion of $G_1$ and $G_2$.
For small $L$, the integrand of \eqref{argclaim2} is then approximately
\begin{equation}
L^{-\nu_1-\nu_2-h_1-h_2-h_3 +h -1}(1-\theta)^{-\nu_1-1}\theta^{-\nu_2-1}\Big\langle\mathcal{O}_1(1)\mathcal{O}_2(\theta)\mathcal{O}_3(0)I\circ\mathcal{O}(0)\Big\rangle_{\rm UHP}.
\end{equation}
The integration over $\theta$ will be finite assuming that the OPE between $\mathcal{O}_2$ and $\mathcal{O}_1$, and between $\mathcal{O}_2$ and $\mathcal{O}_3$, is sufficiently regular; whether this is the case is equivalent to the question of whether the states $\mathcal{O}_1G_1(K)\mathcal{O}_2$ and $\mathcal{O}_2G_2(K)\mathcal{O}_3$ are separately finite, which is not our present concern. Our interest is the convergence of the integration over $L$ towards $L=0$. This will be unproblematic if 
\begin{equation}\nu_1+\nu_2+h_1+h_2+h_3<h.\end{equation}
This is precisely a bound on the leading level of the state $\mathcal{O}_1 G_1(K)\mathcal{O}_2G_2(K)\mathcal{O}_3$ in the dual $\mathcal{L}^-$ expansion.

As a cross check on this argument, consider the state
\begin{equation}\s\bs\s.\end{equation}
For matter sector operators, generally the lowest dimension of a probe state will be $h=0$. Since the leading (and only) level of $\s\bs\s$ is zero, claim \ref{claim2} would imply that the state may be ill-defined. We know that the state is ambiguous in general due to the associativity anomaly. However, the argument given below claim \ref{claim2} does not seem to apply, since \eqref{g1g2} assumes that the operators are separated by elements of the wedge algebra whose leading $\mathcal{L}^-$ level is negative. This can be dealt with by writing $\s\bs\s$ in the form 
\begin{eqnarray}
\s\bs\s &=& -\partial{\s}\frac{1}{1+K}\bs\frac{1}{1+K}\partial\s+(1+K)\s\frac{1}{1+K}\bs\frac{1}{1+K}\partial\s\nonumber\\
&\ & -\partial\s\frac{1}{1+K}\bs\frac{1}{1+K}\s(1+K)+(1+K)\s\frac{1}{1+K}\bs\frac{1}{1+K}\s(1+K).
\end{eqnarray}
Now the argument below claim \ref{claim2} applies to all terms; only the first term can be problematic, since the three boundary condition changing operators form a state whose leading level is zero. Indeed, one finds a divergence from integrating $1/L$ towards $L=0$. However, the state $\s\bs\s$ is not necessarily divergent, only ambiguous. In fact one can check that the integration over $\theta$ actually vanishes towards $L=0$; therefore the first term effectively contains $0\times\infty$, where the ambiguity of $\s\bs\s$ is hidden.

Now let's understand the implications of this for the solution. We want to be certain that 
\begin{equation}\Psi^2\end{equation}
is a well-defined state. For the superstring, the cross terms in $\Psi^2$ which provide the strongest bound on $\nusuper$ arise from the following contributions to the solution as given in \eqref{solutionnoQ}:
\begin{eqnarray}
& & \!\!\!\!\!\!\! \sqrt{H}\s \sqrt{\frac{F}{H}} B\gamma^2\sqrt{\frac{F}{H}}\bs\sqrt{H},\nonumber\\
& & \sqrt{H}\gamma\delta\s BF\gamma\delta\bs\sqrt{H}. \label{an_cont}
\end{eqnarray}
For example, consider the cross terms in $\Psi^2$ where the above contributions are multiplied by
\begin{equation}\sqrt{H}c\frac{1}{H}\s BF\bs\frac{1}{H}c\sqrt{H}.\end{equation}
This gives respectively the states
\begin{eqnarray}
& & \!\!\!\!\!\!\! \sqrt{H}\s \sqrt{\frac{F}{H}} B\gamma^2\sqrt{\frac{F}{H}}\bs\s F\bs\frac{1}{H}c\sqrt{H},\nonumber\\
& & \sqrt{H}\gamma\delta\s BF\gamma\delta\bs \s F\bs\frac{1}{H}c\sqrt{H}.
\end{eqnarray}
Stripping off the ghosts, these states contain the factors
\begin{eqnarray}\s\frac{F}{H}\bs\s &\sim & \s K^{\nu+1}\bs\s,\nonumber\\
\delta\s F\delta\bs\s &\sim & \delta \s K^{\nu} \delta\bs\s,\ \ \ \ (K\to\infty).
\end{eqnarray}
By claim \ref{claim2}, the threefold collision of boundary condition changing operators will be unproblematic if 
\begin{equation}\nusuper<-1 \ \ \ \ { {\rm (no\ triple\ b.c.c.\ operator }\atop {\rm anomalies\ in}\ \eqref{solutionnoQ})}.\end{equation}
Cross terms which do not involve \eqref{an_cont} place a strictly weaker upper bound on $\nusuper$. Therefore, for the superstring $\nusuper<-1$ is sufficient to ensure that \eqref{solutionnoQ} is safe from anomalies due to collisions of boundary condition changing operators. The stronger bound $\nusuper\leq-2$ is sufficient to further guarantee that the formally equivalent expression \eqref{solutionQ} is also safe. Note that the contributions to the solution which give the strongest bound on $\nu$ are only present for the superstring. For the bosonic string, anomalous collisions are absent provided that
\begin{equation}\nubos<0\ \ \ \ { {\rm (no\ triple\ b.c.c.\ operator }\atop {\rm anomalies\ in}\ \eqref{solutionnoQ})}\end{equation}
which safely includes the simple tachyon vacuum.

Let us mention a technical point concerning terms in \eqref{solutionnoQ} involving commutators of $c$ with $\sqrt{F/H}$. Consider for example the contribution
\begin{equation}
\sqrt{H}\s\left[\sqrt{\frac{F}{H}},c\right]\frac{B}{H}\left[c,\sqrt{\frac{F}{H}}\right]\bs\sqrt{H}.\label{dcomc}
\end{equation}
Expanding out the commutator and looking at individual terms, we find that $\s$ and $\bs$ are separated according to
\begin{equation}
\s \frac{F}{H^2}\bs\sim \s K^{\nu+2}\bs,\ \ \ \ (K\to\infty).\label{comexp}
\end{equation}
To avoid anomalous collisions from such terms, it would appear we need $\nu<-2$. On the other hand, since these contributions appear from expanding out a commutator, there may be cancellation of ambiguities. This can be seen more precisely as follows. Let $\sqrt{f/h}(t)$ represent the inverse Laplace transform of $\sqrt{F/H}$, and write 
\begin{eqnarray}
\left[c,\sqrt{\frac{F}{H}}\right] &=& \int_0^\infty dt\, \sqrt{f/h}(t)[c,\Omega^t] \nonumber\\
&=& \int_0^\infty dt\int_0^1d\theta\, \Big(t \sqrt{f/h}(t)\Big)\Omega^{t(1-\theta)}\partial c\Omega^{t\theta}.
\end{eqnarray}
Note the extra factor of $t$ which appears in the integrand. This means, from the point of view of separation of the matter sector boundary condition changing operators, the commutator with $c$ can be seen as equivalent to
\begin{equation}\left[c,\sqrt{\frac{F}{H}}\right] \to -\frac{d}{dK}\sqrt{\frac{F}{H}}.\end{equation}
In particular, in \eqref{dcomc} the boundary condition changing operators are separated as
\begin{equation}\s \left(\frac{d}{dK}\sqrt{\frac{F}{H}}\right)^2\frac{1}{H}\bs\sim \s K^\nu\bs,\ \ \ \ (K\to\infty),\end{equation}
which is significantly more mild than \eqref{comexp}. In this way, terms involving commutators of $c$ with $\sqrt{F/H}$ do not require a stronger bound than $\nusuper<-1$ or $\nubos<0$.

Let us make an important caveat to the above discussion. In \cite{EM} the bosonic string solution was written in two forms:
\begin{eqnarray}
\Psi &=& \frac{1}{\sqrt{1+K}}cB(1+K)c\frac{1}{\sqrt{1+K}}-\frac{1}{\sqrt{1+K}}c(1+K)\s\frac{B}{1+K}\bs(1+K)c\frac{1}{\sqrt{1+K}}\label{sol1}\\
& =& \frac{1}{\sqrt{1+K}}cB(1-\s\bs)(1+K)c\frac{1}{\sqrt{1+K}}-\frac{1}{\sqrt{1+K}}c\partial\s\frac{B}{1+K}\bs(1+K)c\frac{1}{\sqrt{1+K}}.\label{sol2}
\end{eqnarray}
The second form of the solution was useful for computing coefficients in the Fock basis. However, in the second form the computation of $\Psi^2$ will be ambiguous due to associativity anomalies. The origin of the problem is that the first form of the solution, where $\Psi^2$ is well-defined, has been reexpressed as a sum of terms whose star products are individually ambiguous. The fact that this is possible does not reflect poorly on the solution; it is always possible to render a well-defined expression ambiguous by adding and subtracting singular terms. But this raises the possibility that there may be a different way to write the solution for the superstring where ambiguities in $\Psi^2$ disappear, even if $\nusuper\geq-1$. Specifically, the bounds we have derived on the asymptotic behavior of $F(K)$ are sufficient conditions to avoid OPE divergences and associativity anomalies. But they may not be necessary. We leave this question to future work. 

\section{Automorphisms}\label{sec4}

For some purposes it is useful to regard the string field $K$ as having a spectrum consisting of non-negative real numbers. This follows from the observation that the string field 
\begin{equation}\frac{1}{K-\kappa}\end{equation}
is divergent if $\kappa\geq 0$.\footnote{Actually, this state fails to have a well-defined expression as a superposition of wedge states if $\mathrm{Re}(\kappa)\geq 0$. From this point of view, the spectrum of $K$ could be identified with the positive half of the complex plane. However, this may seem unnatural since $K$ is a real string field and ought to have a real spectrum. Furthermore, it is not completely clear that the state is actually divergent for complex $\kappa$ with positive real part. A finite expression for the state in the Virasoro basis can be given following \cite{exotic}. The prescription is to write the wedge state $\Omega^\alpha$ in the Virasoro basis, and for each appearance of $\frac{1}{(\alpha+1)^h}$ in the expansion coefficients we replace
\begin{equation}\frac{1}{(1+\alpha)^h}\to \frac{1}{(h-1)!}\int_0^\infty d K\, K^{h-1}e^{-K}\frac{1}{K-\kappa}.\end{equation}
The integral is finite as long as $\kappa$ is not zero or positive, which supports the idea that the spectrum should be non-negative reals. However, more work is needed to verify that this prescription gives an adequate definition of the state. For the purposes of the present discussion we leave these subtleties to the side, and proceed under the assumption that the spectrum of $K$ is real and non-negative.} Consider the connected component\footnote{There has been interesting discussion of diffeomorphisms of the spectrum of $K$ which are not homotopic to the identity, generated by the transformation $K\to1/K$ \cite{Hata}. The resulting automorphisms are singular, and we will not consider them.} of the diffeomorphism group on the spectrum 
\begin{equation}{\rm Diff}_0(\mathbb{R}_{\geq 0}).\end{equation}
This automatically defines a group of automorphisms of the algebra of wedge states, defined through
\begin{equation}
\phi\circ G(K) = G(\phi(K)),\ \ \ \ \phi\in{\rm Diff}_0(\mathbb{R}_{\geq 0}).
\end{equation}
More surprisingly, it is possible to generalize this into an automorphism group of the $KBc$ subalgebra~\cite{erler-simple}. Applications of this symmetry have been discussed in \cite{dual-L,Hata,erler-simple,MNT,Hata2,Masuda,TT-KOS,Ishibashi}. Here we show that the automorphisms can be further extended to act on boundary condition changing operators. Applying such automorphisms to the solution \eqref{EM} turns out to be equivalent to changing the choice of tachyon vacuum. Therefore the solutions discussed in this work can be related through diffeomorphism of the spectrum of $K$.

The solution lives in a subalgebra given by multiplying generators
\begin{equation}K,\ \ B,\ \ c,\ \ \gamma^2,\ \ \s,\ \ \bs,\ \ Q\s,\ \ Q\bs.\end{equation}
This is a graded differential associative $*$-algebra. An automorphism of such an algebra should satisfy 
\begin{eqnarray}
{\rm gh}(\phi\circ A)&=&{\rm gh}(A),\\
\phi\circ (QA) &=& Q(\phi\circ A),\\
\phi\circ(AB) &=& (\phi\circ A)(\phi\circ B), \\
\phi\circ(A^\ddag) &=& (\phi\circ A)^\ddag,
\end{eqnarray}
where ``${\rm gh}$'' denotes ghost number and $\ddag$ denotes reality conjugation of the string field. An important part of realizing the automorphism group is understanding what relations the generators of the algebra should satisfy. This is not completely trivial, since the relevant collection of identities is actually a proper subset of those satisfied by $K,B,c,\gamma^2,\s,\bs,Q\s,Q\bs$ as defined in the conventional way by operator insertions on the identity string field, as assumed in section \ref{sec2}. We postulate the following algebraic relations,
\begin{eqnarray}
&\ &\!\!\!\!\!\!\!\!\! Bc+cB = 1,\ \ \ B^2=c^2=0,\ \ \ [K,B] = 0;\nonumber\\
&\ &\!\!\!\!\!\!\!\!\! [B,\gamma^2] = [c,\gamma^2] = 0;\nonumber\\
&\ &\!\!\!\!\!\!\!\!\! [B,\s]=[B,\bs] = 0,\ \ \ [c,\s]=[c,\bs]=0;\nonumber\\
&\ &\!\!\!\!\!\!\!\!\! \bs\s = 1;\nonumber\\
&\ &\!\!\!\!\!\!\!\!\! [B,Q\s]=[K,\s],\ \ \ [B,Q\bs] = [K,\bs];\label{alg_rel}
\end{eqnarray}
the following differential relations,
\begin{eqnarray}
&\ &\!\!\!\!\!\!\!\!\! QB = K,\ \ \ QK = 0,\ \ \ Qc = cKc -\gamma^2\nonumber\\
&\ &\!\!\!\!\!\!\!\!\! Q\gamma^2 = cK\gamma^2-\gamma^2 Kc;\label{BRST_rel}
\end{eqnarray}
and the following properties under conjugation: 
\begin{eqnarray}
&\ &\!\!\!\!\!\!\!\!\!  K^\ddag = K,\ \ \ B^\ddag = B,\ \ \ c^\ddag = c;\nonumber\\
&\ &\!\!\!\!\!\!\!\!\!  (\gamma^2)^\ddag = \gamma^2;\nonumber\\
&\ &\!\!\!\!\!\!\!\!\! \s^\ddag = \bs,\ \ \ \bs^\ddag = \s;\nonumber\\
&\ &\!\!\!\!\!\!\!\!\! (Q\s)^\ddag = -Q\bs,\ \ \ (Q\bs)^\ddag = -Q\s.\label{real_rel}
\end{eqnarray}
The conventional understanding of the generators as operator insertions on the identity string field results in an infinite number of additional (and less important) relations, referred to as ``auxiliary identities" in \cite{dual-L}. A notable example of such a relation is
\begin{equation}[\s,\gamma^2] = 0.\label{aux}\end{equation} 
This identity is absent from \eqref{alg_rel}, and henceforth we assume that it does not hold. 

Given $\phi\in{\rm Diff}_0(\mathbb{R}_{\geq 0})$ we have a corresponding automorphism of the above subalgebra defined by
\begin{eqnarray}
\widehat{K} \ =\ \phi\circ K & = & \phi(K),\\
\widehat{B}\ =\ \phi\circ B & =&  B\frac{\phi(K)}{K},\\
\widehat{c} \ = \ \ \, \phi\circ c &= &cB\frac{K}{\phi(K)}c,\\
\widehat{\gamma^2}\  = \ \phi\circ \gamma^2 &=& cB\frac{K}{\phi(K)}\gamma^2+\gamma^2\frac{K}{\phi(K)}Bc,\\
\widehat{\s}\  = \ \ \, \phi\circ\s &=&  cB\sqrt{\frac{K}{\phi(K)}}\s\sqrt{\frac{\phi(K)}{K}}+\sqrt{\frac{\phi(K)}{K}}\s\sqrt{\frac{K}{\phi(K)}}Bc,\\
\widehat{\bs} \ = \ \ \, \phi\circ\bs &=&  cB\sqrt{\frac{K}{\phi(K)}}\bs\sqrt{\frac{\phi(K)}{K}}+\sqrt{\frac{\phi(K)}{K}}\bs\sqrt{\frac{K}{\phi(K)}}Bc.
\label{aut}
\end{eqnarray}
One can check that the transformed generators (denoted with hat) satisfy all relations \eqref{alg_rel}-\eqref{real_rel}. However, they do not satisfy auxiliary identities, such as \eqref{aux}. The automorphism is characterized by two states in the wedge algebra
\begin{equation}\frac{\phi(K)}{K},\ \ \ \frac{K}{\phi(K)}.\label{aut_pair}\end{equation}
These states look potentially singular at $K=0$. To ensure that they are well-defined, we work with the connected component of the diffeomorphism group, and further assume that the diffeomorphisms are (at least) once differentiable at $K=0$. This implies
\begin{equation}\phi(0) = 0,\ \ \ \phi'(0)>0,\label{diff_ass}\end{equation}
which is sufficient to ensure that the automorphism is regular at $K=0$. One particularly simple type of diffeomorphism is a scale transformation of the spectrum of $K$
\begin{equation}\phi(K) = \alpha K,\ \ \ \alpha>0.\end{equation}
This leads to 
\begin{equation}
\phi\circ K = \alpha K,\ \ \ \phi\circ B = \alpha B,\ \ \ \phi\circ c = \frac{1}{\alpha}c,\ \ \ \phi\circ\gamma^2 = \frac{1}{\alpha}\gamma^2,\ \ \ \phi\circ\s= \s,\ \ \ \phi\circ\bs = \bs.
\end{equation}
In this case the automorphism is equivalent to the well-known midpoint-preserving reparameterization generated by $\mathcal{L}^-$. 

Now we can understand how the automorphisms act on the solution \eqref{EM}. If the solution is characterized by some $F(K)$, one can show that the tachyon vacuum \eqref{tvs} and the boundary condition changing fields \eqref{SbSQ} transform as
\begin{eqnarray}\phi\circ \Psi_{\rm tv}[F] &=& \Psi_{\rm tv}[\phi\circ F],\nonumber\\
\phi\circ\S[F] &=& \S[\phi\circ F],\nonumber\\
\phi\circ\bS[F] &=& \bS[\phi\circ F],\end{eqnarray}
where the dependence on $F$ is explicitly shown in the brackets. This immediately implies that the solution itself transforms as
\begin{equation}
\phi\circ\Psi[F] = \Psi[\phi\circ F].
\end{equation}
Therefore the automorphism simply changes the choice of $F$ characterizing the solution, or equivalently, the choice of tachyon vacuum. Note that \eqref{diff_ass} implies that the automorphism preserves the conditions \eqref{F_cond} on the choice of $F$. Particularly relevant for present purposes is understanding how diffeomorphisms affect the collision of boundary condition changing operators inside the solution. If we wish to transform from $F$ whose leading level is $\nu<0$ in the dual $\mathcal{L}^-$ expansion, to $F'$ whose leading level is $\nu'<0$, the leading level of $\phi(K)$ should be
\begin{equation}\frac{\nu'}{\nu}>0.\end{equation}
The level is positive, so $\phi(K)$ necessarily diverges as $K\to\infty$; this is consistent with the assumption that $\phi$ is in the connected component of the diffeomorphism group of $\mathbb{R}_{\geq 0}$. Note that if we want to produce a more regular solution, so that $\nu'$ is more negative than $\nu$, the diffeomorphism necessarily grows more than linearly for large $K$. In a sense, to produce a more regular solution we need to push the spectrum of $K$ out to infinity. This means that the more regular the desired solution, the more singular the state $\phi(K)$ must be from the perspective of the identity string field \cite{dual-L}. An extreme example of this is the diffeomorphism relating the simple tachyon vacuum to Schnabl's solution:
\begin{equation}\phi(K) = \Omega^{-1}-1.\label{sch-rep}\end{equation}
The inverse wedge state $\Omega^{-1}$ is so singular that it does not even have a well-defined Fock space expansion. To avoid this kind of problem one can  impose conditions on the asymptotic behavior of elements in ${\rm Diff}_0(\mathbb{R}_{\geq 0})$ towards $K=\infty$, at the cost of limiting the range of solutions that can be related by automorphism, though we will not do so here. If a choice of $F$ can be related to $\frac{1}{1+K}$ through diffeomorphism, the above analysis implies that it is formally possible to express the solution in the same form as the solution based on the simple tachyon vacuum: 
\begin{eqnarray}
\Psi[F] \!\!\!&=&\!\!\! \frac{1}{\sqrt{1+\widehat{K}}}\Big(\widehat{c}+Q(\widehat{B}\widehat{c})\Big)\frac{1}{\sqrt{1+\widehat{K}}}-\frac{1}{\sqrt{1+\widehat{K}}}\widehat{\s}\Big(\widehat{c}+Q(\widehat{B}\widehat{c})\Big)\widehat{\bs}\frac{1}{\sqrt{1+\widehat{K}}}\nonumber\\
&\ &+\frac{1}{\sqrt{1+\widehat{K}}}Q\widehat{\s} \frac{\widehat{B}}{1+\widehat{K}}Q\widehat{\bs}\frac{1}{\sqrt{1+\widehat{K}}}-\frac{1}{\sqrt{1+\widehat{K}}}Q\widehat{\s} \widehat{B}\widehat{c}\widehat{\bs}\frac{1}{\sqrt{1+\widehat{K}}}+\frac{1}{\sqrt{1+\widehat{K}}}\widehat{\s}\widehat{c}\widehat{B}Q\widehat{\bs}\frac{1}{\sqrt{1+\widehat{K}}},\nonumber\\
\end{eqnarray}
with the appropriately transformed generators. One must be careful however, since this expression inherits many of the problems with boundary condition changing operator collisions which appear for the simple tachyon vacuum. For example, the term
\begin{equation}\frac{1}{\sqrt{1+\widehat{K}}}\widehat{\s}\widehat{c}\widehat{\bs}\frac{1}{\sqrt{1+\widehat{K}}} = \sqrt{F}cB\sqrt{\frac{KF}{1-F}}\s\bs\sqrt{\frac{KF}{1-F}}c\sqrt{F}\end{equation}
contains the undesirable product $\s\bs$. We know that the solution is well-defined from the point of view of boundary condition changing operator collisions if the leading level of $F$ is $-2$ or lower. Therefore such singular terms must formally cancel. 

It is not always possible to relate two $F$s by diffeomorphism of the spectrum of $K$. It will only be possible if they share the same number of extrema and values at the extrema. This implies that tachyon vacuum and boundary condition changing operator solutions can be partitioned into equivalence classes under diffeomorphism symmetry. Since any change of $F$ subject to \eqref{F_cond} amounts to a gauge transformation, diffeomorphism of the spectrum of $K$ gives a finer notion of equivalence than is provided by gauge symmetry. Whether this finer notion of equivalence has some deeper meaning is not clear.

\section{Conclusion}

Having taken care of boundary condition changing operator collisions, we now have an infinite class of well-defined solutions of cubic superstring field theory which can describe any time-independent background. However, cubic superstring field theory is a somewhat handicapped framework; what we really want is the analogue of this solution in the context of the nonpolynomial Wess-Zumino-Witten-like superstring field theory. The difficulty here is that there is only one known analytic solution for the tachyon vacuum \cite{supervac}, and it seems difficult to modify it so as to provide additional separation between $\s$ and~$\bs$. This is most likely a technical problem. The tachyon vacuum of \cite{supervac} is an expression for the group element $e^\Phi$ of the Wess-Zumino-Witten-like action, and the leading level of $\Phi$ in the dual $\mathcal{L}^-$ level expansion is $-1/2$. To regulate boundary condition changing operator collisions we need the leading level to be strictly less than $-1/2$, and such solutions always seem to require infinite sums over correlation functions containing increasing numbers of superghost insertions for each component field of the Fock space expansion. It is not known how to understand the convergence of these sums. However, it is not difficult to find formal algebraic expressions for such solutions, and provided convergence issues can be brought under control, we would have the desired generalization of \cite{EM} to the Wess-Zumino-Witten-like theory. Whether this is the most useful way to proceed is not clear to us, but as a general matter it would be desirable to have analytic control of a wider class of tachyon vacuum solutions in the nonpolynomial framework. 

One fascinating complication of the Wess-Zumino-Witten-like formulation is that tachyon vacuum solutions are not guaranteed to be universal, and for stable D-brane systems may not even exist. The straightforward generalization of \cite{EM} therefore only seems to apply to D-branes with vanishing topological charge. By contrast, in cubic superstring field theory all open string backgrounds have a universal solution for the tachyon vacuum, even if tachyons are absent from the spectrum \cite{Erler-tv-cubic}. Therefore a complete generalization of \cite{EM} to the nonpolynomial framework requires coming to terms with how D-brane charges are realized in the string field algebra, which opens the way to a whole class of new and interesting questions. We hope that the present work is a useful step in this direction.

\newpage

\noindent{\bf Acknowledgements}
\vspace{.5cm}

\noindent The authors thank the Yukawa Institute for Theoretical Physics at Kyoto University
for kind invitation to  the workshop YITP-T-18-04 ``New Frontiers in String Theory 2018"
which provided a stimulating environment where this research was discussed. 
The work of C.M.  is partially supported by the MIUR PRIN
Contract 2015MP2CX4 Non-perturbative Aspects Of Gauge Theories And Strings. The work of T.E. is supported by ERDF and M\v{S}MT (Project CoGraDS -CZ.02.1.01/0.0/0.0/15\_ 003/0000437) and the GA{\v C}R project 18-07776S and RVO: 67985840.



\begin{thebibliography}{99}
 
 
 \bibitem{EM}
  T.~Erler and C.~Maccaferri,
  ``String Field Theory Solution for Any Open String Background,''
  JHEP {\bf 1410} (2014) 029
  [arXiv:1406.3021 [hep-th]].
  
\bibitem{KOS}   
  M.~Kiermaier, Y.~Okawa and P.~Soler,
  ``Solutions from boundary condition changing operators in open string field theory,''
  JHEP {\bf 1103}, 122 (2011)
  [arXiv:1009.6185 [hep-th]].

 
 \bibitem{Ishibashi:2018ynb}
  N.~Ishibashi, I.~Kishimoto, T.~Masuda and T.~Takahashi,
  ``Vector profile and gauge invariant observables of string field theory solutions for constant magnetic field background,''
  JHEP {\bf 1805} (2018) 144
  [arXiv:1804.01284 [hep-th]].

\bibitem{Maccaferri:2015cha}
  C.~Maccaferri and M.~Schnabl,
  ``Large BCFT moduli in open string field theory,''
  JHEP {\bf 1508} (2015) 149
  [arXiv:1506.03723 [hep-th]].


 \bibitem{Ishibashi:2016xak}
  N.~Ishibashi, I.~Kishimoto and T.~Takahashi,
  ``String field theory solution corresponding to constant background magnetic field,''
  PTEP {\bf 2017} (2017) no.1,  013B06
  [arXiv:1610.05911 [hep-th]].
 
  \bibitem{Kishimoto:2014yea}
  I.~Kishimoto, T.~Masuda, T.~Takahashi and S.~Takemoto,
  ``Open String Fields as Matrices,''
  PTEP {\bf 2015} (2015) no.3,  033B05
  [arXiv:1412.4855 [hep-th]].

\bibitem{PTY}

  C.~R.~Preitschopf, C.~B.~Thorn and S.~A.~Yost,
  ``Superstring Field Theory,''
  Nucl.\ Phys.\ B {\bf 337}, 363 (1990).


\bibitem{Russians}

  I.~Y.~Arefeva, P.~B.~Medvedev and A.~P.~Zubarev,
  ``New Representation for String Field Solves the Consistency Problem for Open Superstring Field Theory,''
  Nucl.\ Phys.\ B {\bf 341}, 464 (1990).


\bibitem{Berkovits}

  N.~Berkovits,
  ``SuperPoincare invariant superstring field theory,''
  Nucl.\ Phys.\ B {\bf 450}, 90 (1995)
  Erratum: [Nucl.\ Phys.\ B {\bf 459}, 439 (1996)]
  [hep-th/9503099].


\bibitem{simple-tv}
  T.~Erler and M.~Schnabl,
  ``A Simple Analytic Solution for Tachyon Condensation,''
  JHEP {\bf 0910} (2009) 066
  [arXiv:0906.0979 [hep-th]].



\bibitem{Schnabl}
  M.~Schnabl,
  ``Analytic solution for tachyon condensation in open string field theory,''
  Adv.\ Theor.\ Math.\ Phys.\  {\bf 10} (2006) no.4,  433
  [hep-th/0511286].
  
\bibitem{Okawa}

 Y.~Okawa,
  ``Comments on Schnabl's analytic solution for tachyon condensation in Witten's open string field theory,''
  JHEP {\bf 0604}, 055 (2006)
  [hep-th/0603159].
  
\bibitem{dual-L}
  T.~Erler,
  ``The Identity String Field and the Sliver Frame Level Expansion,''
  JHEP {\bf 1211} (2012) 150
  [arXiv:1208.6287 [hep-th]].
  
  
\bibitem{exotic}
 
   T.~Erler,
  ``Exotic Universal Solutions in Cubic Superstring Field Theory,''
  JHEP {\bf 1104}, 107 (2011)
  [arXiv:1009.1865 [hep-th]].
  

\bibitem{Erler-tv-cubic}
  T.~Erler,
  ``Tachyon Vacuum in Cubic Superstring Field Theory,''
  JHEP {\bf 0801} (2008) 013
  [arXiv:0707.4591 [hep-th]].
  
\bibitem{Jokel}
  M.~Jokel,
  ``Real Tachyon Vacuum Solution without Square Roots,''
  arXiv:1704.02391 [hep-th].

\bibitem{SSFII}

  T.~Erler,
  ``Split String Formalism and the Closed String Vacuum, II,''
  JHEP {\bf 0705}, 084 (2007)
  [hep-th/0612050].


\bibitem{MurataSchnabl}

  M.~Murata and M.~Schnabl,
  ``Multibrane Solutions in Open String Field Theory,''
  JHEP {\bf 1207}, 063 (2012)
  [arXiv:1112.0591 [hep-th]].


  
\bibitem{tensor}

  M.~R.~Gaberdiel and B.~Zwiebach,
  ``Tensor constructions of open string theories. 1: Foundations,''
  Nucl.\ Phys.\ B {\bf 505}, 569 (1997)
  [hep-th/9705038].
  
 \bibitem{Hata}
 
   H.~Hata and T.~Kojita,
  ``Singularities in K-space and Multi-brane Solutions in Cubic String Field Theory,''
  JHEP {\bf 1302}, 065 (2013)
  [arXiv:1209.4406 [hep-th]].

\bibitem{erler-simple}
  T.~Erler,
  ``A simple analytic solution for tachyon condensation,''
  Theor.\ Math.\ Phys.\  {\bf 163} (2010) 705
   [Teor.\ Mat.\ Fiz.\  {\bf 163} (2010) 366].

\bibitem{MNT}

  T.~Masuda, T.~Noumi and D.~Takahashi,
  ``Constraints on a class of classical solutions in open string field theory,''
  JHEP {\bf 1210}, 113 (2012)
  [arXiv:1207.6220 [hep-th]].
  
\bibitem{Hata2}

  H.~Hata and T.~Kojita,
  ``Inversion Symmetry of Gravitational Coupling in Cubic String Field Theory,''
  JHEP {\bf 1312}, 019 (2013)
  [arXiv:1307.6636 [hep-th]].

\bibitem{Masuda}  
    T.~Masuda,
  ``Comments on new multiple-brane solutions based on Hata-Kojita duality in open string field theory,''
  JHEP {\bf 1405}, 021 (2014)
  [arXiv:1211.2649 [hep-th]].

  
 \bibitem{TT-KOS}
  C.~Maccaferri,
  ``A simple solution for marginal deformations in open string field theory,''
  JHEP {\bf 1405} (2014) 004
  doi:10.1007/JHEP05(2014)004
  [arXiv:1402.3546 [hep-th]].
  
  \bibitem{Ishibashi}
  
    N.~Ishibashi,
  ``Comments on Takahashi-Tanimoto’s scalar solution,''
  JHEP {\bf 1502}, 168 (2015)
  [arXiv:1408.6319 [hep-th]].

\bibitem{supervac}

  T.~Erler,
  ``Analytic solution for tachyon condensation in Berkovits` open superstring field theory,''
  JHEP {\bf 1311}, 007 (2013)
  [arXiv:1308.4400 [hep-th]].


 \end{thebibliography}
\end{document}